# Control of Spontaneous Orientation Polarization in Organic Semiconductors: The Role of Molecular Structure and Film Growth Conditions

### Albin Cakaj, Markus Schmid, Alexander Hofmann, and Wolfgang Brütting

wolfgang.bruetting@physik.uni-augsburg.de
Institute of Physics, University of Augsburg, 86135 Augsburg, Germany

Keywords: molecular orientation, giant surface potential, organic light-emitting diodes

**ABSTRACT**

*Spontaneous orientation polarization (SOP) occurs when molecules with a finite permanent dipole moment are grown as thin films by physical vapor deposition and their alignment is such that a net non-zero polarization remains. We discuss how SOP in organic semiconductors can be controlled by the design of molecules as well as the film growth conditions and discuss its relevance in organic light-emitting devices.*

## 1 Introduction

Spontaneous orientation polarization (SOP) occurs when molecules with a finite permanent dipole moment (PDM) are grown as thin films by physical vapor deposition and their alignment is such that a net non-zero polarization remains. It manifests itself by a so-called giant surface potential (GSP) – typically of the order of some 10-100 mV/nm (or equivalently $10^7$-$10^8$ V/m of electric field strength) – or the presence of surface/interface charges of a few mC/m$^2$.

SOP in organic semiconductors was "discovered" more than 20 years ago [1,2]. However, only several years later, it was realized that both observations – the occurrence of interfacial charges as well as the giant surface potential – have SOP as their common origin [3].

Since then, a broad range of organic semiconductor materials has been studied, mainly by Kelvin probe on single-layer thin films or by impedance spectroscopy on hetero-layer device structures consisting of one nonpolar layer and a second layer showing SOP [4]. Over the years, one has gathered a substantial data set that provides qualitative insights into the occurrence of SOP in organic semiconductors and its magnitude in relation to the PDM of the studied molecules (see Fig. 1).

The key findings can be summarized as:
- SOP occurs only in vapor-deposited films, while solution-cast films of the same molecular materials show no preferential alignment of their dipole moments.
- Most of the studied materials exhibit positive GSP slopes, i.e. they align with the positive partial charge of their PDM vectors pointing away from the substrate surface.
- The degree of dipole alignment Λ, i.e. the magnitude of the GSP slope in relation to the situation with fully aligned PDMs along the surface normal, is rather small; typically, between 4 and 12% only.

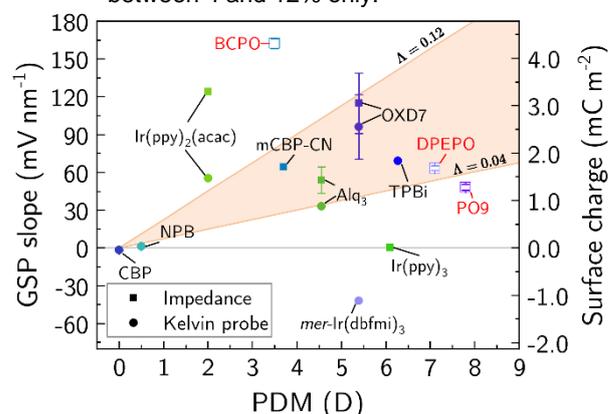

*Fig. 1: Overview of the SOP of polar organic semiconductors studied by us. GSP slope measured by Kelvin probe and surface charge taken from impedance spectroscopy are plotted vs. the electric dipole moment of the molecules. For the conversion between one and the other a dielectric constant of 3 was assumed. (Reprinted with permission from ACS Appl. Mater. Interfaces 2023, 15, 47, 54721–54731. Copyright 2023, The Authors.)*

Most of the research on SOP so far has focused on single-component neat films of polar organic molecules, although dilution of a polar species in a non-polar matrix, i.e. dipolar doping, has been studied in some cases [5,6].

Meanwhile, new material designs have allowed to tune and control both the magnitude and the sign of the SOP over a wide range. Specifically, Tanaka et al. have published an innovative material design concept that independently allows tuning the alignment of molecules by chemical asymmetry of the molecular backbone as well as the orientation of their PDMs by proper functionalization. In that way, record-high GSP values of ±150 mV/nm could be achieved [7].

Furthermore, the concept of surface equilibration was introduced by the Ediger group to explain the anisotropic alignment of organic molecules in thin films grown by physical vapor deposition. Specifically, the substrate temperature ($T_S$) during film growth in relation to the glass transition temperature of the bulk material ($T_g$) was identified as key parameter for controlling the film structure of such anisotropic molecular glasses [8].

Here we present molecular design rules to control SOP in organic semiconductor thin films, as well as the effect of film growth conditions on the magnitude of SOP. Further, we discuss the implications of SOP in the context of organic light-emitting devices (OLEDs).

## 2 SOP and its Control by Molecular Design and Film Growth Conditions

As a first example, we have studied TPBi as a prototypical electron transport material in OLEDs (see Fig. 2). TPBi has a PDM of 6 – 8 D (depending on the conformer; see below) and a glass transition temperature of 122°C. The molecules is often visualized with a pyramidal structure, where the central phenyl ring and the benzimidazol groups of the three arms form the base plane, while the PDM is orthogonal to this plane and, thus, points to the tip of the pyramid (Fig. 2b). However, it was recently shown by Wang et al. that this highly symmetric structure is just one possible isomer (the so-called C3 isomer with a PDM of 7.6 D), while the lower symmetry C1 isomer, where one of the three arms is rotated by 180°, has a PDM of only 6 D (Fig. 2a) [9]. Most importantly however, in this case the PDM vector is no longer orthogonal to the base plane. Due to their slightly different energetics, the C1 isomer is naturally more abundant with 83%, while there is only 17% of the C3 isomer. This leads to the typically observed small SOP values of TPBi films (~60 – 70 mV/nm). Introducing p-ethyl substitutions at the phenyl rings of the three arms, Wang et al. could achieve 100% isomer purity and, thus, boost the GSP by 82%.

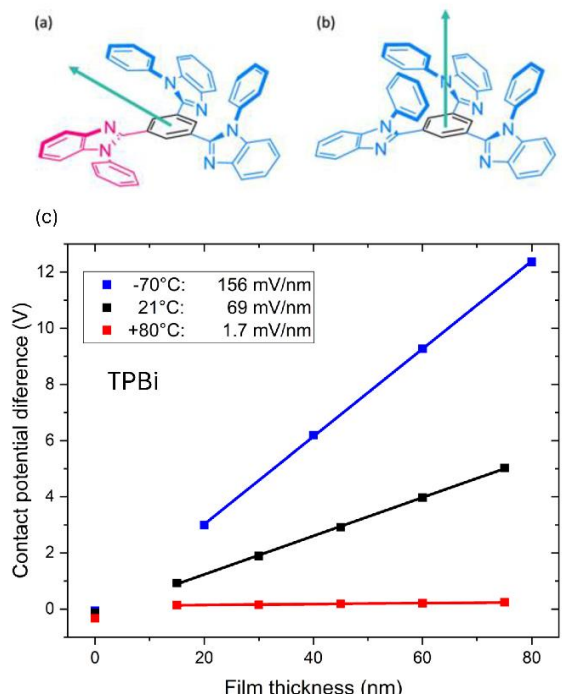

Fig. 2: Structures of the two TPBi isomers with a) C1 symmetry and b) C3 symmetry. (Reprinted with permission from ACS Appl. Mater. Interfaces 2022, 14, 16, 18773–18781. Copyright 2022, American Chemical Society.) c) Measured contact potential difference vs. film thickness at three different substrate temperatures, from where the given GSP slopes are calculated.

Another way to control the SOP of evaporated films is a variation of the substrate temperature used for film growth. As shown in Fig. 2c, in this way we can boost the GSP to more than 150 mV/nm, if the substrate is cooled to -70°C, or vice versa, if the substrate is heated to +80°C, the GSP vanishes. This behavior is in good agreement with the above-mentioned surface equilibration model, where film anisotropies are expected to disappear at a value of $T_S/T_g > 0.85$.

Another class of organic semiconductors are so-called phosphine oxides, having a strongly polar P=O bond in their structure (see Fig. 3), which are frequently used as host or electron transport materials in thermally activated delayed fluorescence or phosphorescent OLEDs. Among them are two molecules with just one P=O bond (BCPO & POPy2) having low PDM of about 3.5 D, and the other with two P=O groups (DPEPO & PO9), where the (average) PDM is about 7 – 8 D. Surprisingly, however, their GSPs are not correlated to the PDM magnitude. By contrast, BCPO and POPy2 have more than twice of the SOP strength than the other two, which means that, regarding their lower PDM, their degree of PDM alignment is 4 – 6 times higher (see Table 1).

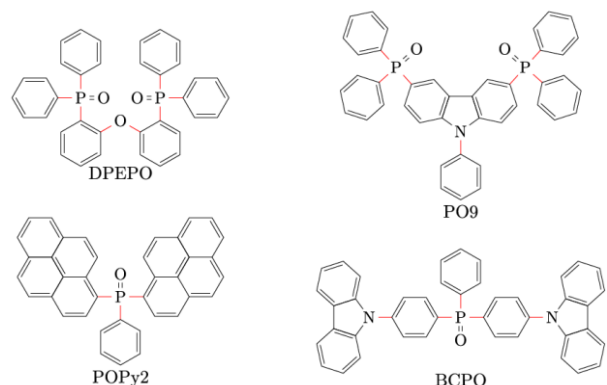

Fig. 3: Chemical structures of the investigated phosphine oxide containing molecules. (Reprinted with permission from ACS Appl. Mater. Interfaces 2023, 15, 47, 54721–54731. Copyright 2023, The Authors.)

| Material | PDM (D) | $T_g$ (°C) | GSP (mV/nm) | Alignment Λ (%) |
|---|---|---|---|---|
| DPEPO | 7.1 | 93 | 63.3 | **5.5** |
| PO9 | 7.8 | 122 | 48.2 | **4.5** |
| POPy2 | 3.6 | NA | 129 | **21** |
| BCPO | 3.5 | 137 | 162 | **32** |

Tab. 1: Properties of the investigated phosphine oxide containing molecules and films grown by physical vapor deposition with the substrate kept at room temperature.

As explained in detail in Ref. [10], this is due to the possibility of forming many different conformers with varying dihedral angles between the rotatable groups adjacent to the phosphor atom in these molecules. And, if there are more than one P=O group, this causes partial compensation of their PDMs and prevents preferential alignment of them. We also want to note that the GSP of BCPO is with 162 mV/nm among the highest values reported so far. Furthermore, we have shown that by using substrate cooling to -70°C we can almost double the value close to 300 mV/nm (see Fig. 4), and the overall dependence on $T_S$ again follows the surface equilibration model [10].

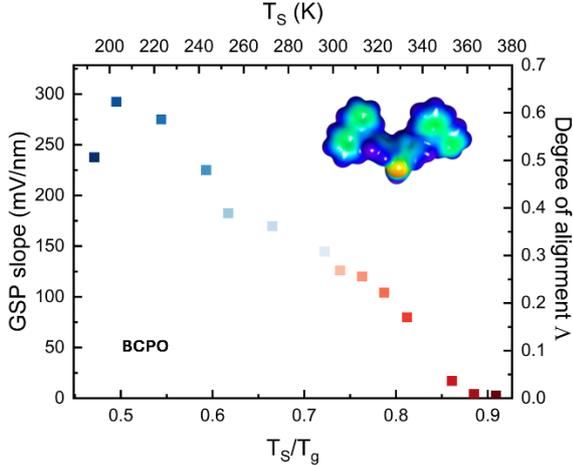

*Fig. 4: GSP slopes of BCPO films grown at different substrate temperature $T_S$. and the corresponding degree of PDM alignment Λ. The inset shows the electrostatic surface potential of the molecule, where the oxygen atom (orange) with its negative potential aligns preferentially pointing downward to the substrate. (Reprinted with permission from ACS Appl. Mater. Interfaces 2023, 15, 47, 54721–54731. Copyright 2023, The Authors.).*

Another important point concerns the actual orientation distribution of the PDMs in a film grown by physical vapor deposition. One has to be aware that the alignment parameter Λ (derived from the GSP slope or the surface charge density) is an average value over all the angles enclosed between the PDMs of the molecules and the substrate normal (see Fig. 5a). More precisely, Λ is proportional to the first moment of a yet unknown orientation distribution function $f(\theta)$. So far, there are only simulation studies of the film growth process yielding insights into the shape of $f(\theta)$ for different polar molecules exhibiting SOP [11], but it is often considered "difficult to experimentally determine the ratio of the face-up and face-down orientations in the films" [9].

Here we make use of the fact that the orientation distribution function $f(\theta)$ may be expanded into Legendre polynomials $P_l(\cos\theta)$ such that it can be written as:

$$f(\theta) = \sum_l \frac{2l+1}{2} \langle P_l \rangle \cdot P_l(\cos\theta).$$

Therein, each of the Legendre polynomials represents a specific moment of the distribution function. Combining different experiments, it is now possible to obtain several of these coefficients $\langle P_l \rangle$ so that an approximation for $f(\theta)$ is possible. Per definition, $\langle P_0 \rangle = 1$, $\langle P_1 \rangle$ is the SOP order parameter Λ, and $\langle P_{2,4,6} \rangle$ can be obtained from nuclear magnetic resonance experiments [12].

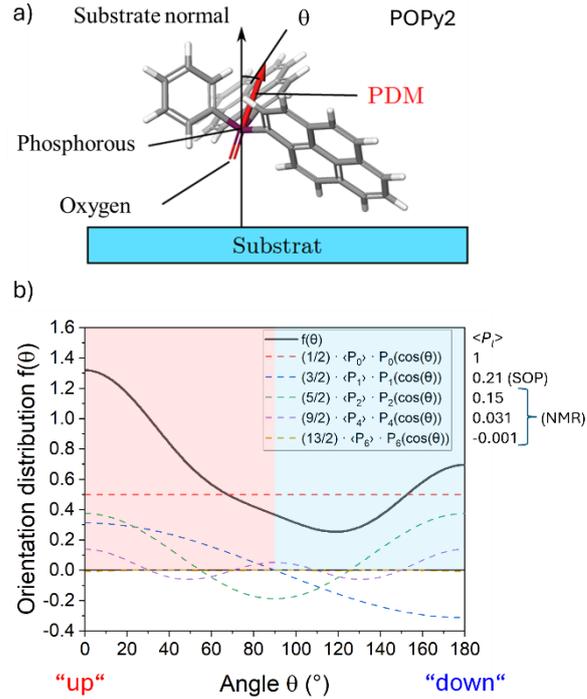

*Fig. 5: a) Preferential alignment of a POPy2 molecule on the substrate surface with the PDM vector enclosing an angle $\theta$ to the substrate normal. b) Orientation distribution $f(\theta)$ of the PDM vectors of POPy2 in films grown by physical vapor deposition with the substrate kept at room temperature.*

As shown in Fig. 5b, this yields a rough idea of the shape of $f(\theta)$, with most of the molecular PDMs of POPy2 pointing upward on the substrate, just a few lying flat in the film plane and some pointing downward. This means that the non-ideal alignment parameter of Λ=0.21 is a consequence of a considerable fraction of misaligned PDMs pointing downward, but not due to PDMs lying mostly in plane with just a small portion of out-of-plane aligned PDMs pointing upward, as has been observed in many of the simulation results [11].

### 3 Discussion and Conclusions

It is well established that SOP is an inherent feature of many organic semiconductor thin films, if they are made by physical vapor deposition of polar molecules. Nevertheless, the question of whether (or not) it is beneficial for application in OLEDs has been controversial [13]. Early on, it was shown that the presence of a positive GSP in electron transporting materials is favorable for electron injection, even though it usually is accompanied by a slight increase in the turn-on voltage [14]. However, recently it has been found out that the concomitant sub-turn-on hole accumulation in the (nonpolar) hole transport layer causes exciton quenching that is undesired in OLEDs [15]. But the possibility to control SOP, through the methods presented above, and also by dipolar doping of otherwise nonpolar materials, opens a wide range tunability. E.g. Noguchi et al. have demonstrated that it is indeed possible to design OLED stacks that have identical, non-vanishing SOP in all of their layers so that charge accumulation and exciton quenching does no longer occur [16]. And there may be other requirements on specific device functions, where the presence of a built-in electric field in an organic semiconductor film is beneficial. One such example is vibration power generators using self-aligned SOP materials [17]. We may expect even more to come in the future [18-20].